\documentclass[prl,floatfix,reprint]{revtex4-2}

\usepackage{amsmath}
\usepackage{amssymb}
\usepackage{graphicx}
\usepackage{hyperref}
\usepackage{color}
\usepackage{bm}

\usepackage{xcolor}
\definecolor{codegreen}{rgb}{0,0.6,0}
\definecolor{backcolour}{rgb}{0.95,0.95,0.92}

\usepackage{listings}
\lstdefinestyle{mystyle}{
    backgroundcolor=\color{backcolour},   
    commentstyle=\color{codegreen},
    keywordstyle=\color{magenta},
    basicstyle=\ttfamily\footnotesize,
}
\begin{document}

\title{Quantum Synchronization of Fock States}

\author{Fabian Hassler, David Scheer, Samah Saquaque, Steven Kim}
\affiliation{Institute for Quantum Information, RWTH Aachen University, 52056 Aachen, Germany}
\date{May 2026}
\begin{abstract}
Synchronization, a ubiquitous phenomenon in classical systems, has recently been extended to the quantum domain.
Here, we show quantum synchronization of a bosonic mode exhibiting a Fock state-like limit cycle, manifesting as a steady state with a negative Wigner function.
We demonstrate that this non-classical state can be phase-locked to an external drive, achieving synchronization within an Arnold tongue regime.
We argue that synchronization is a dynamical property and fundamentally tied to the suppression of phase slips, which we show to occur with exponentially decreasing probability.
We introduce a novel method to extract the phase slip rate from the Lindblad time evolution of the system.
This work opens new avenues for understanding and manipulating non-classical synchronization dynamics.
\end{abstract}
\maketitle

\textit{Introduction}.---%
First observed by Huygens in 1665 \cite{huygens}, synchronization is a universal phenomenon found across diverse physical and biological systems \cite{pikovsky}.
While synchronization manifests in fascinating ways, such as the synchronous flickering of fireflies \cite{fireflies}, it also serves as a vital tool in technology, exemplified by the Josephson voltage standard, which relies on the synchronization of Josephson oscillations \cite{voltage}.
Current research actively explores synchronizing Bloch oscillations, with the potential to establish a novel current standard \cite{shaikhaidarov:22,crescini:23,kaap:24}.

The extension of synchronization to the quantum domain has recently garnered significant attention, with explorations across platforms including spins and few-level systems \cite{ bergli:20, qubit}, circuit QED \cite{ankerhold:25, paaske:26}, optomechanical arrays \cite{fazio:13, marquardt:14, marquardt:16, lorch:17}, and trapped ions \cite{lesanovsky:15, armour:20, kehrer:25}---often leveraging quantum analogs of Stuart-Landau or van der Pol oscillators as a theoretical foundation \cite{lee:13,bruder:14, bruder:16,chia:25}.
This approach has proven valuable for stabilizing coherent radiation and squeezed states \cite{scheer:24, squeezing:25}.

However, existing quantum synchronization studies are often accurately described by semi-classical models, and demonstrating true quantum signatures has proved challenging \cite{bruder:16}.
Furthermore, existing measures of synchronization, based on probability distributions \cite{bruder:16}, correlations \cite{armour:15}, or information \cite{fazio:15, fazio:20}, typically focus on stationary states, making it difficult to definitively distinguish a synchronized state from a phase-dependent, non-synchronized one.

Since synchronization fundamentally represents a dynamical process \cite{pikovsky}, a dynamical measure is essential.
Ideally, synchronization stabilizes a specific oscillation phase $\phi_0$.
Driven-dissipative systems are typically required to observe synchronization, and these systems are inherently subject to noise.
While noise can be suppressed in classical systems, quantum systems are always affected by some level of noise.
Beyond the inherent uncertainty in $\phi_0$, noise can induce rapid $2\pi$ phase slips.
However, in classical synchronization, such phase slips are exponentially suppressed---a defining characteristic that has largely been overlooked in quantum systems, with a notable exception being Ref.~\cite{paaske}.
Therefore, we adopt this dynamical characteristic as a key benchmark for synchronization in the quantum regime.
Combining this dynamical feature with a non-classical state defines what we term \textit{quantum synchronization}.

In this Letter, we present the synchronization of a Fock state---a prime example of a non-classical state---to an external drive.
We begin by describing a superconducting circuit capable of stabilizing a Fock state in the stationary state, as previously demonstrated in Ref.~\cite{clerk:16}.
Given the non-classical nature of the Fock state, a typical semi-classical approach is not warranted.
Therefore, we develop a quantum theory enabling the study of phase dynamics in the quantum setting.
We utilize this formalism to derive a Fokker-Planck equation for the phase variable that characterizes the phase-locking process.
To establish the presence of synchronization, we investigate phase diffusion within the quantum regime, with a particular focus on phase-slips.
For this purpose, we introduce a novel method allowing access to phase dynamics from the generator of the time evolution of the system.

\begin{figure}[b]
	\centering
	\includegraphics{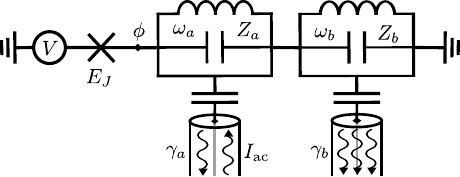}
	\caption{%
Superconducting circuit in which the quantum synchronization of a Fock state to an external drive can be observed.
It consists of a Josephson junction, with Josephson energy $E_J$, that is put in series with two LC-resonators with frequencies $\omega_j$ and characteristic impedances $Z_{j}$, $j\in\{a,b\}$.
The DC voltage-bias $V$ is tuned to the value $2eV =\hbar(\omega_a+\omega_b)$ such that a tunneling Cooper pair can resonantly excite one photon in each resonator.
The resonators are capacitively coupled to transmission lines, described by the damping rates $\gamma_j$.
Through the transmission line, an external drive $I_\mathrm{ac}$ can be applied to resonator $a$ which allows the study of quantum synchronization to an external drive.
}\label{fig:setup}
\end{figure}

\textit{The model}.---%
To study quantum synchronization of Fock states, we require a model that can stabilize them.
The setup is displayed in Fig.~\ref{fig:setup}.
It consists of a Josephson junction with Josephson energy $E_J$, that is biased by a DC voltage $V$, and is put in series with two single-mode resonators.
Each resonator, $j\in\{a,b\}$, is characterized by a resonance frequency $\omega_j$ and a characteristic impedance $Z_{j}$.
The voltage across the resonators $(\hbar/e)\dot \phi$ is given by the time-derivative of the node flux $\phi = \sqrt{\alpha_a}(a+a^\dag) + \sqrt{\alpha_b}(b+b^\dag)$.
Here, $\alpha_j = (4 \pi e^2/h)Z_j$ and $a,b$ are bosonic ladder operators that fulfill the canonical commutator $[a,a^\dag]=[b, b^\dag]=1$.
Without loss, this system can be described by the Hamiltonian \cite{armour:15, peugeot:21, clerk:16}
\begin{equation}
H = \hbar\omega_a a^\dag a + \hbar \omega_b b^\dag b - E_J \cos\bigl(2eVt/\hbar - \phi\bigr).
\end{equation}

When the DC bias-voltage is tuned such that $2eV = \hbar(\omega_a+\omega_b)$, the system exhibits a parametric resonance. 
This leads to self-sustained oscillations, which can be employed to stabilize a Fock state~\cite{clerk:16}.
To study this, we move to a rotating-frame via the unitary transformation $U(t) = \exp(i\omega_a a^\dag a t+i\omega_b b^\dag b t)$ and perform a rotating-wave approximation.
We obtain the effective Hamiltonian
\begin{equation*}
H_\mathrm{RWA} = \tilde E_J a^\dag b^\dag \, {}_1\!F_1(-a^\dag a, 2, \alpha_a)\,{}_1\!F_1(-b^\dag b, 2, \alpha_b)+ \mathrm{H.c.}
\end{equation*}
with $\tilde E_J=\sqrt{\alpha_a\alpha_b}E_J e^{-(\alpha_a + \alpha_b)/2}/2$ and the confluent hypergeometric function ${}_1F_1(a,b,c)$ \cite{abramowitz}.

Including damping, the time evolution of the density matrix of the composite system can be described by a Lindblad equation with
\begin{equation}\label{eq:Lindblad}
\dot \rho = \mathcal{L}\rho = -\frac{i}{\hbar}[H_\mathrm{RWA},\rho] + \sum_{j\in\{a,b\}} \mathcal{D}[L_j]\rho
\end{equation}
with the Liouvillian $\mathcal{L}$ and dissipator $\mathcal{D}[L]\rho = L\rho L^\dag - (L^\dag L \rho + \rho L^\dag L)/2$.
The jump operators $L_j = \sqrt{\gamma_j}\, j$ describe single photon loss at rate $\gamma_j$.
Typically, the temperature is small $k_B T\ll \hbar \omega_j$ such that thermal effects are negligible.

When resonator $b$ is strongly damped with $\gamma_b \gg \gamma_a$, we can assume that the $b$-mode is always in its ground state.
This allows to trace out the contribution of resonator $b$ which leads to an effective description for resonator $a$.
We obtain the Lindblad equation
\begin{equation}\label{eq:Lindblad_fock}
\dot \rho = \gamma_a\mathcal{D}[a]\rho + \epsilon\, \mathcal{D}[a^\dag \, {}_1\!F_1(-a^\dag a, 2, \alpha_a)]\rho
\end{equation}
with $\epsilon = 2 \tilde E_J^2/\hbar^2\gamma_b$.
Note that this model is invariant under the transformation $a \mapsto a e^{i\vartheta}$ and, thus, has a $U(1)$-symmetry.

\begin{figure}[tb]
	\centering
	\includegraphics[]{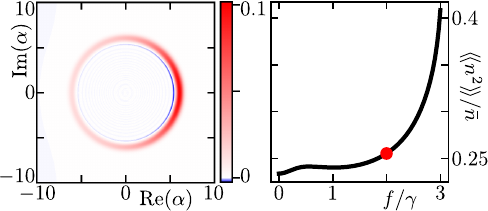}
	\caption{%
(a) Wigner function of resonator $a$ with parameters $\epsilon=20\gamma_a$, $n_0 = 50$, $\Delta=0$, $\theta=0$, and $f=2\gamma_a$.
We see that the external drive injects its phase into the system via the localization of the Wigner function at $\phi=\theta=0$.
Importantly, the negativity persists and oscillations can be observed inside the limit-cycle, resembling the features of a Fock state.
(b) Ratio of the variance $\langle \! \langle n^2 \rangle\!\rangle$ to the average photon number $\bar{n}$.
For increasing driving strength $f$, the ratio increases but remains smaller than $1$. The red dot corresponds to $f=2\gamma_a$ for which the Wigner function is shown in (a).
}\label{fig:wigner}
\end{figure}

The system stabilizes a Fock state due to competing single photon loss ($\gamma_a$) and nonlinear single photon gain ($\epsilon$).
The gain process, described by ${}_1F_1(-a^\dag a, 2, \alpha_a)=\sum_n L_n^{(1)}(\alpha_a)/(n+1) |n\rangle\langle n|$, prohibits transitions to states with $n>n_0$ when $\alpha_a$ is a zero of the Laguerre polynomial $L_n^{(1)}(\alpha_a)$.
Strong pumping ($\epsilon \to \infty$) drives the system into the Fock state $n_0$, similar to the photon blockade \cite{koch:06, vaneph:18, zhou:25} utilized for single photon sources \cite{rolland:19}.
Stabilizing high-$n_0$ Fock states requires $\alpha_a \approx 3.67/n_0$, corresponding to a resonator impedance $Z_a \approx 7.56 \mathrm{k\Omega}/n_0$.
For finite, but large $\epsilon$ the stationary distribution $P_n$ in Fock space is a Gaussian with mean $\bar{n} \approx n_0\bigl(1-\sqrt{\gamma_a/\epsilon J^2}\bigr)$ and variance $\langle\!\langle n^2 \rangle \!\rangle \approx n_0\sqrt{\gamma_a/4J^2\epsilon}$ ($J \approx -0.40$), see \cite{suppl}.
The narrow Gaussian,  with a width scaling like $\langle \!\langle n^2 \rangle \!\rangle /\bar{n} \propto \epsilon^{-1/2}$, approximates a Fock state, resulting in a negative Wigner function (Fig.~\ref{fig:wigner})---a signature of a non-classical state essential for quantum synchronization.
Note that the negativity increases with $n_0$ \cite{kenfack:04}.

\textit{Quantum phase dynamics}.---%
To study synchronization, we want to study the dynamics of the phase of the system.
Typically, a semi-classical approach is chosen, where $\langle a \rangle = \alpha = re^{-i\phi}$ is decomposed into a radial component $r$ and a phase $\phi$ \cite{bruder:26}.
However, the state of our system is not classical, \textit{i.e.}, the Wigner function is negative.
Thus, a semi-classical approach is not justified. 
To study the dynamics of the phase in this case, we require a quantum theory, which we formulate in the following.

The starting point is the phase distribution $P(\phi,t) = \langle \phi|\rho(t) |\phi\rangle$ with the phase states $|\phi\rangle = \sum_{n=0}^\infty e^{-in\phi}|n\rangle$.
For a general density matrix $\rho(t) = \sum_{n,l} \rho_{n+l,n}(t) |n+l\rangle\langle n|$, the phase distribution can be written as $P(\phi, t) = \sum_{l\in \mathbb{Z}} P_l(t)e^{il\phi}$ where $P_l(t) = \sum_{n=0}^\infty  \rho_{n+l,n}(t)$\cite{pires:18}.
Note that as we sum over the photon number $n$, we are left with a marginal (positive) probability distribution for the phase.
Because the phase $\phi \in [0,2\pi)$ is compact, $l$ is an integer representing its conjugate angular momentum.

For a $U(1)$-symmetric system, as described by Eq.~(\ref{eq:Lindblad_fock}), each $l$-sector decouples, allowing independent study of the dynamics within each sector.
Without an external drive, the system's stationary state resides entirely within the $l=0$-sector, meaning the corresponding density matrix is diagonal and the system is rotationally invariant.
We now investigate the coupling of the system to an external coherent drive---also known as entrainment---which breaks the $U(1)$-symmetry and consequently couples the different $l$-sectors.

\textit{Entrainment}.---%
In our setup, the external drive can be realized by an AC current-bias $I_\mathrm{ac}(t) = I_0\sin(\theta-\omega_d t )$.
In the rotating frame, this leads to the Hamiltonian 
\begin{equation}\label{eq:H_drive}
H_d=\hbar \Delta a^\dag a + i\hbar f(a^\dag e^{-i\theta} - a e^{i\theta})
\end{equation} 
with a small detuning $|\Delta| = |\omega_a-\omega_d| \ll \omega_a$ and driving strength $f=I_0\sqrt{Z_a/8\hbar}$.
This adds the contribution $-i[H_d,\rho]/\hbar$ to the Lindblad equation~\eqref{eq:Lindblad_fock}.

As mentioned above, for $f=0$, the stationary state of the system is close to a Fock state.
In particular, the Wigner function is negative and reveals the quantum limit cycle with radius $r_0 \approx \sqrt{\bar{n}}$.
For finite driving, $f>0$, the $U(1)$-symmetry is broken and the phase of the system $\phi$ locks to the phase of the external drive $\theta$.
Importantly, the negativity persists such that the system remains in a non-classical state, see Fig.~\ref{fig:wigner}.

To gain an understanding of the phase-locking, we want to derive an equation of motion for $P(\phi, t)$.
Note that the relation
\begin{equation}
\dot P(\phi, t) = \langle \phi|\dot \rho(t) |\phi\rangle = \langle \phi|\mathcal{L} \rho(t) |\phi\rangle  = \sum_{l\in \mathbb{Z}} \dot P_l(t)e^{il\phi}
\end{equation}
holds.
Thus, $\dot P(\phi, t)$ can be obtained by transforming the Lindblad equation to an equation of motion for $P_l(t)$ and applying the Fourier transform.
This leads to a Fokker-Planck equation for $P(\phi,t)$ given by
\begin{equation}\label{eq:FP}
\dot P =  -\partial_\phi \biggl[\Delta - \frac{f}{\sqrt{\bar{n}}} \sin(\phi-\theta) \biggr] P + D\partial_\phi^2 P
\end{equation}
with $D = [\gamma_a+\epsilon \, {}_1\!F_1(-\bar{n}, 2, \alpha_a)]/8\bar{n} \approx \sqrt{\gamma_a \epsilon}/8n_0$.
In the derivation of this, it is necessary that the probability distribution $P_n$ is close to a Fock state with  $\bar n \gg 1$. Details on the derivation can be found in the End Matter.

The first term, linear in $\partial_\phi$, is called drift and is due to the motion in an effective potential $U(\phi)$.
The second term corresponds to diffusion due to noise.
Due to the $U(1)$-symmetric limit cycle, the free diffusion constant $D$ is inversely proportional to the photon number, $D \propto \gamma_a/\bar{n}$ \cite{scully, schawlow}. 
This is also known as the Schawlow-Townes limit.
The Fokker-Planck equation is equivalent to the noisy Adler equation $\dot\phi = -\partial_\phi U(\phi) + \xi(t)$ with the tilted washboard potential $U(\phi) = -\Delta \phi - \frac{f}{\sqrt{\bar{n}}}\cos(\phi-\theta)$.
The Gaussian white noise $\xi(t)$ has zero mean and variance $\langle \!
\langle \xi(t)\xi(t')\rangle \!
\rangle = 2D\delta(t-t')$.
For $f>|\Delta|\sqrt{\bar{n}}$ \footnote{Note that this condition is necessary but not sufficient as we point out throughout this work.}, the potential develops minima and the phase stabilizes at $\phi_0 = \theta + \arcsin(\Delta \sqrt{\bar{n}}/f)$.
This locking, however, is not perfect because of the fluctuations.
They lead to an uncertainty in the phase given by $\langle (\phi-\phi_0)^2 \rangle = D\sqrt{\bar{n}}/\sqrt{f^2-\Delta^2\bar{n}}$.

This locking of the phase in the stationary state is often taken as a characteristic of quantum synchronization. 
However, the external drive \eqref{eq:H_drive} always forces its phase onto any system, independent of the fact if a limit cycle is present or not.
As such, it only measures how close the state of the system is to a coherent state.
In particular, the external drive also acts as a displacement and injects the photon number $n_f = f^2/(\Delta^2+\gamma_a^2/4)$ into the system \cite{portugal:23}.
Thus, only weak driving with $n_f \ll \bar n$ should be considered. 
For $n_f \gtrsim \bar n$, the state of the system is indistinguishable from a coherent state.
In the End Matter, we point out that by only taking the stationary state into account, that even a simple coherent state can show signatures of `synchronization'.
Due to these reasons, instead, we employ a dynamical property of synchronization as a synchronization measure.

\textit{Phase diffusion}.---%
In order to understand synchronization, we have to look at the extended phase $\varphi$ given by $\varphi(t)-\varphi(0) = \int_0^t \dot \phi(\tau) d\tau$.
Without the external drive, $f=0$, the phase diffuses freely with $\langle \! \langle [\varphi(t)-\varphi(0)]^2 \rangle \! \rangle = 2Dt$.
In the synchronized regime, the phase is trapped at $\phi_0$ mod $2\pi$.
However, due to noise there are phase-slips which change $\varphi$ by $\pm 2\pi$ in a rapid event. 
We denote the correspondingly directional phase-slip rates as $\Gamma_\pm$.

These rare phase-slips resemble a noise-induced escape from a metastable state, also called the Kramers' problem \cite{kramers:40}.
From this, it is expected that the escape rates are exponentially small with $\Gamma_\pm \propto \exp(-\Delta U_\pm/D)$ where $\Delta U_\pm$ is the directional barrier height, which can be obtained from the tilted washboard potential $U(\phi)$.
Without detuning, $\Delta=0$, the rates are equal and scale according to $\Gamma_\pm \propto \exp(-8 f\sqrt{n_0/\gamma_a \epsilon})$.

In the stationary state, the phase-slips lead to two phenomena \cite{pikovsky}. 
A drift of the phase given by
\begin{equation}\label{eq:drift}
\langle \varphi(t) - \varphi(0) \rangle \simeq 2\pi(\Gamma_+ - \Gamma_-)t, \qquad (t\to \infty)
\end{equation}
which vanishes for $\Delta=0$ and is exponentially small else, and phase diffusion with
\begin{equation}\label{eq:diffusion}
\langle\!\langle [\varphi(t)-\varphi(0)]^2 \rangle \!\rangle \simeq 4\pi^2(\Gamma_+ + \Gamma_-)t,
\end{equation}
with $\varphi(0)=\phi_0 =  \theta + \arcsin(\Delta \sqrt{\bar{n}}/f) $.
Compared to the free case, synchronization thus leads to an exponential suppression of phase diffusion; this suppression is the hallmark signature of classical synchronization in the presence of noise \cite{pikovsky}.
As quantum systems are never noiseless, we adopt this dynamical definition also for the quantum case.

Detecting synchronization requires a phase $\varphi$ defined in the extended zone scheme to have access to the phase-slip rate.
However, as noted above, the angular momentum $l$ is an integer, which corresponds to the compact phase $\phi$.
We would like to obtain a generating function of the extended phase $\varphi(t)$ to obtain the drift and diffusion, see Eq.~\eqref{eq:drift} and~\eqref{eq:diffusion}.
For this purpose, we define a phase operator $\hat \varphi$ as conjugate to the number operator with $[\hat \varphi, a^\dag a]=i$.
This operator induces shifts in Fock states, such that $\exp(i q\hat \varphi) |n\rangle = |n+q\rangle$ for $q \in \mathbb{R}$.
Since $q$ is not restricted to integer values, $\hat \varphi$ describes the phase within an extended zone scheme.
Although the definition of a phase operator is formally problematic due to the lower bound on photon number \cite{phase}, this definition is sufficient for our purposes, as the limit cycle confines the photon number away from the vacuum.

The cumulants of the extended phase can be obtained from 
\begin{align}\label{eq:generating_func}
	\langle \! \langle [ \varphi(t)-\varphi(0)]^m \rangle \! \rangle = (-i \partial_q)^m \ln \langle e^{iq[\varphi(t)-\varphi(0)]}\rangle\big|_{q=0}
\end{align}
where $q$ serves as a counting field for the extended phase, providing access to phase-slips.
The moment generating function can be obtained from 
\begin{equation*}
\langle e^{iq[\varphi(t)-\varphi(0)]}\rangle = \operatorname{Tr}\bigl[e^{iq\hat\varphi} e^{\mathcal{L}t} e^{-iq\hat\varphi}\rho(0)\bigr] = \operatorname{Tr}\bigl[e^{\mathcal{L}(q) t}\rho(0)\bigr].
\end{equation*}
The deformed Liouvillian $\mathcal{L}(q)$ is obtained by defining superoperators $O_+$ acting on the density matrix as $O_+ \rho \equiv O\rho$.
The shift operators $e^{\pm i q \hat \varphi}$ always act to the left of the density matrix, resulting in $\mathcal{L}(q) = e^{iq\hat\varphi_+} \mathcal{L} e^{-iq\hat\varphi_+}$.
Therefore, to obtain $\mathcal{L}(q)$, the ladder operators ($a$ and $a^\dag$) acting on the left of the density operator must be replaced by their deformed counterparts, $a(q) = e^{iq\hat\varphi} a e^{-iq\hat\varphi}$, satisfying $a(q) |n\rangle = \sqrt{n-q}|n-1\rangle$, see \cite{suppl}.

Since the time evolution is generated by the Liouvillian $\mathcal{L}$, it is also possible to obtain the generating function from its eigenvalues \cite{derrida:98, bruderer:14, kim:23}.
We are specifically interested in the properties of the non-equilibrium steady state, defined by $\dot \rho = \mathcal{L}\rho=0$.
Thus, the eigenvalue $\lambda_0(q)$ that is adiabatically connected to the stationary state with $\lambda_0(q=0)=0$ is of importance.
For long times, it can be shown that $\ln \langle e^{iq[\varphi(t)-\varphi(0)]}\rangle \sim \lambda_0(q) t$.

\begin{figure}[tb]
	\centering
	\includegraphics[]{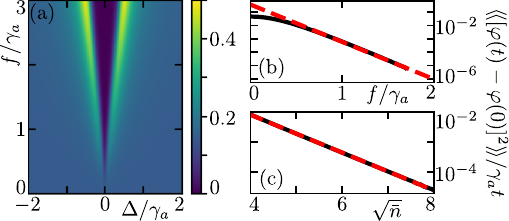}
	\caption{%
		Phase diffusion for different parameters; $\epsilon = 20\gamma_a$ and $n_0 = 50$ besides in (c) where $n_0$ is varied. (a) Diffusion as a function of external driving strength $f$ and detuning $\Delta$. We see that the system goes from free phase diffusion (unsynchronized) to exponentially suppressed phase diffusion (synchronized) for small $\Delta$ or large $f$. The diagram reveals a so-called Arnold tongue, a benchmark for synchronization. Panels (b) and (c) reveal the exponential suppression. Black lines correspond to a numerical evaluation while the red dashed lines are fits to $\ln \Gamma_\pm \propto -f\sqrt{\bar n}$ as expected from the Kramers' result. In (b), $\Delta=0$, while in (c), $\Delta=0, f=\gamma_a$, and $n_0 \in [30,80]$ leading to different $\bar{n}$. The parameters are limited due to numerical precision because of the exponentially small quantities involved.
}\label{fig:diffusion}
\end{figure}

The result for the phase diffusion obtained from Eq.~\eqref{eq:generating_func} can be seen in Fig.~\ref{fig:diffusion}.
In Fig.~\ref{fig:diffusion}(a), we keep the Fock state fixed and show the diffusion constant as a function of the detuning $\Delta$ and the coupling strength $f$.
The plot resembles an Arnold tongue and shows the phase transition from the unsynchronized to the synchronized phase.
In the unsynchronized regime, the diffusion is free.
For increasing coupling $f$, the diffusion first increases until it goes into the synchronized regime where the diffusion is exponentially suppressed.
The initial increase in diffusion stems from the dynamical phase transition, leading to enhanced fluctuations at the critical point \cite{reimann:02}.
Figures~{\ref{fig:diffusion}}(b) and (c) confirm the exponential scaling of the escape rates $\Gamma_\pm$.

\textit{Conclusion}.---%
To conclude, we have studied the synchronization of a Fock state, a prime example of a state that exhibits a non-classical limit cycle, to an external drive.
As the Fock state is highly non-classical, a semi-classical approach was not viable. 
For this reason, we have formulated a quantum theory based on phase states, which allows to study the dynamics of the phase. 
Using this formalism, we have derived a Fokker-Planck equation that is equivalent to the Adler equation in the precense of noise, the paradigm for the study of classical synchronization.

As quantum systems are always subject to noise, we have emphasized that stationary properties are not sufficient to characterize synchronization.
Thus, we adopted the exponential suppression of phase diffusion, the hallmark signature of classical synchronization with noise, as the synchronization measure to the quantum regime.
This requires access to the phase in an extended zone scheme.
Based on generating functions, we were able to formalize a theory for the counting of phase-slips which also allows for a efficient numerical implementation.

Our results open avenues for several future directions.
Experimentally, the findings presented here can be measured using existing setups; in particular, the photon blockade scheme needed to stabilize a Fock state has been previously demonstrated for single-photon source generation~\cite{rolland:19}.
Theoretically, the methods and formalism introduced offer a new perspective on the existing literature, which has largely focused on the stationary state.
Furthermore, given that in our system the suppression of the phase-slip rates require substantial photon numbers, a more detailed investigation of synchronization in few-level systems and the `deep quantum limit' is warranted.
Finally, exploring the mutual synchronization of Fock states is an interesting question for furture work.

\textit{Acknowledgements}.---%
F.H. would like to thank C.\ Bruder for insightful discussions that initiated the exploration of Fock state synchronization.
All authors acknowledge fruitful discussions with C.\ Bruder, H.\ Christiansen, T.\ Nadolny, and J.\ Paaske.
This work was supported by the Deutsche Forschungsgemeinschaft (DFG) under Grant No.\ HA 7084/8--1.

\clearpage
\onecolumngrid\clearpage\appendix
\setcounter{equation}{0}\renewcommand\theequation{A\arabic{equation}}
\setcounter{figure}{0}\renewcommand\thefigure{A\arabic{figure}}\renewcommand\theHfigure\thefigure
\setcounter{table}{0}\renewcommand\thetable{A\Roman{table}}\renewcommand\theHtable\thetable
\setcounter{page}{1}
\makeatletter\let@environment{thebibliography}{NAT@thebibliography}\makeatother
\renewcommand{\bibnumfmt}[1]{[A#1]}
\renewcommand{\citenumfont}[1]{A#1}
\section*{End Matter}
\begin{quote}
In the first part of this End Matter, we will demonstrate that the stationary state alone is not a suitable measure of synchronization.
We show that even a simple coherent state can exhibit a feature resembling an `Arnold tongue' when analyzed solely based on the stationary probability distribution.
This highlights the necessity of considering the dynamical properties of synchronization, as we do in the main text.
Furthermore, based on our formalism of quantum phase dynamics introduced in the main text, we derive the Fokker-Planck equation~(\ref{eq:FP}) in the second part of this appendix.
\end{quote}

\subsection{Stationary States vs.\ Dynamics}

Here, we demonstrate that a coherent state, driven externally, can produce a phase distribution that mimics an `Arnold tongue' when considering only the stationary probability distribution.
Such a coherent state can be realized by externally driving a damped harmonic oscillator, a process described by the Lindblad equation
$\dot \rho = -\frac{i}{\hbar}[H_d, \rho] + \gamma \,\mathcal{D}[a]\rho,$
where $H_d =\hbar \Delta a^\dag a + \hbar f(a e^{i\theta} + a^\dag e^{-i\theta})$ represents the driving Hamiltonian.
The stationary state of this system is a coherent state $\rho_{0} =|\alpha\rangle\langle\alpha|$, where $|\alpha \rangle = e^{-|\alpha|^2/2}\sum_{n=0}^\infty \alpha^n |n\rangle /\sqrt{n!}$ and $\alpha = fe^{-i\theta}/(\Delta-i\gamma/2)$.

A common approach for characterizing synchronization is to consider the maximum of the phase distribution $P_0(\phi)=P(\phi, t\rightarrow \infty)$.
This distribution is defined as $P_0(\phi) = \langle \phi |\rho_0|\phi \rangle$ with the phase states $|\phi\rangle = \sum_{n=0}^\infty e^{-i n \phi}|n\rangle$.
For a coherent state, the distribution is given by
\begin{equation}
P_0(\phi) = e^{-|\alpha|^2}\left|\sum_{n=0}^\infty \frac{(\alpha e^{i\phi})^n}{\sqrt{n!}}\right|^2 = \begin{cases}
1 + 2|\alpha|\cos(\phi-\phi_0), &  |\alpha| \ll 1 \\
\sqrt{8\pi|\alpha|^2} \exp\bigl[ -2|\alpha|^2(\phi-\phi_0)\bigr],   &  |\alpha| \gg 1
\end{cases}
\end{equation}
with $\alpha = |\alpha|e^{-i\phi_0}$.
Notably, this distribution exhibits phase localization at the phase of the coherent drive, $\phi=\phi_0$.
Consequently, $\mathrm{max}[P_0(\phi)]$ is often used as a synchronization measure.
However, this metric only reflects the \textit{localization} of the phase at a particular value $\phi_0$.

\begin{figure}[h]
\centering
\includegraphics[scale=0.9]{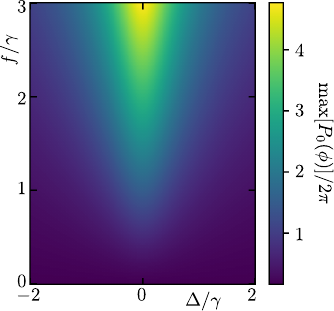}
\caption{Phase localization measure $\mathrm{max}[P_0(\phi)]$ as a function of external driving strength $f$ and detuning $\Delta$.  This plot exhibits a feature resembling an `Arnold tongue', but it solely characterizes the efficiency of the external drive and is not a reliable indicator of synchronization.
}
\label{fig:coherent}
\end{figure}

As we demonstrate here, even a simple externally driven harmonic oscillator exhibits this characteristic, indicating it is not a property unique to synchronization.
Specifically, Fig.~\ref{fig:coherent} displays $\mathrm{max}[P_0(\phi)]$ as a function of the external driving strength $f$ and detuning $\Delta$.
In both limiting cases of $|\alpha|$, $\mathrm{max}[P_0(\phi)]$ varies linearly with $|\alpha|=f/\sqrt{\Delta^2+\gamma^2/4}$.
This contrasts with our synchronization measure, which reveals an exponential dependence.
The observed phase localization, previously interpreted as a synchronization indicator, appears as an `Arnold tongue'-like feature.
However, in this context, it merely reflects the efficiency of the external drive, characterized by strong driving and small detuning, and does not genuinely signify synchronization.
Therefore, a dynamical measure of synchronization, as emphasized in the main text, is essential.
Moreover, it is essential that the drive is always kept small such that the drive is a perturbation of the limit-cycle and not the other way round.
\newpage
\subsection{Fokker-Planck equation}
Here, we show how to obtain the Fokker-Planck equation~\eqref{eq:FP} for the marginal probability distribution $P(\phi, t)$.
In particular, we make use of
\begin{equation}
\dot P(\phi, t) = \langle \phi|\dot \rho(t) |\phi\rangle = \langle \phi|\mathcal{L} \rho(t) |\phi\rangle  = \sum_{l\in \mathbb{Z}} \dot P_l(t)e^{il\phi}.
\end{equation}
As $P(\phi,t)$ and $P_l(t)$ are connected via a Fourier transform, $P_l(t) = \int d\phi\, P(\phi, t)e^{-il\phi}/2\pi$, we want to obtain an equation of motion for $P_l(t)$ and use the equivalence $lP_l(t) \leftrightarrow -i\partial_\phi P(\phi, t)$.
In general, $\dot P_l(t)$ contains all orders of $l$, however, as we are interested in phase diffusion, it suffices to expand up to second order in $l$.

The Liouvillian contains three parts, a single photon loss, nonlinear photon gain, and the Hamiltonian $H_d =\hbar \Delta a^\dag a + i\hbar f(a^\dag e^{-i\theta} - a e^{i\theta})$.
In the following, we examine how each of them impact the time evolution of $P_l(t)$.
The main approximation will be that the probability distribution $P_n = \langle n|\rho|n\rangle$ peaks at $\bar{n} \gg 1$, as it is the case for our system.
This property will also holds true for $l \neq 0$, which become important for $f\neq0$.
However, as long as the system is not described by a coherent state, $n_f \ll \bar{n}$ (see main text), only small $l$ will be relevant.

For the single photon loss process, we obtain
\begin{equation}
\langle \phi|\mathcal{D}[a]\rho|\phi\rangle = \sum_{n,l}\bigl[\sqrt{(n+l+1)(n+1)}\rho_{n+l+1, n+1} - (n+l/2) \rho_{n+l, n} \bigr]e^{il\phi}
\end{equation}
Expanding for large photon numbers, $n \gg l$, leads to 
\begin{align}
\langle \phi|\mathcal{D}[a]\rho|\phi\rangle &\approx \sum_{n,l}\Bigl[\bigl(n+l/2-l^2/8n\bigr)\rho_{n+l+1, n+1} - (n+l/2) \rho_{n+l, n} \Bigr]e^{il\phi} \\
&=\sum_{l}\Bigl[-l\rho_{l,0}/2-\sum_{n=1}^\infty l^2 \rho_{n+l, n}/8n\Bigr]e^{il\phi} \approx - \sum_l l^2 P_l(t) e^{il\phi}/8\bar n \,,  
\end{align}
where we have assumed that $P_n$ is peaked at $\bar n \gg 1$.
This assumptions lets us neglect the `boundary' term $\rho_{l,0}\approx 0$ and insert the average photon number.
Without the property that $P_n$ is peaked at $\bar n \gg 1$, our approach of deriving the Adler equation fails.
Similarly, it can be shown that the contribution of the nonlinear gain is given by $\bigl[-l^2 \,{}_1\!F_1(-\bar{n}, 2, \alpha_a) /8\bar{n}\bigr]P_l(t)$, again neglecting boundary terms.

Next, we investigate the impact of the Hamiltonian $H_d$, which contributes via $-i[H_d,\rho]/\hbar$.
We obtain, for large photon numbers,
\begin{equation}
-i\langle \phi |[H_d,\rho] |\phi\rangle/\hbar \approx -i\sum_l l\Bigl[ \Delta  - \frac{f}{\sqrt{\bar{n}}} \sin(\phi-\theta)\Bigr] P_l(t) e^{il\phi}
\end{equation}
which is linear in $l$.
We obtain the dynamics of $P(\phi, t)$ via a Fourier transform. 
In particular, we employ the relations $l P_l(t) \leftrightarrow -i\partial_\phi P(\phi)$ and $l^2 P_l(t) \leftrightarrow -\partial_\phi^2 P(\phi, t)$, which leads to 
\begin{equation}
\dot P(\phi, t) = -\partial_\phi \Bigl[\Delta  - \frac{f}{\sqrt{\bar{n}}} \sin(\phi-\theta) \Bigr] P(\phi, t) + \frac{\gamma_a + \epsilon \,{}_1\!F_1(-\bar{n}, 2, \alpha_a)}{8\bar{n} }\partial_\phi^2 P(\phi, t),
\end{equation}
as given in the main text.
 
\newpage
\section*{Supplemental material}
\begin{quote}
In this supplemental material, we show how the setup given in the main text is able to stabilize a Fock state. We show that the stationary state $P_n$ is a Gaussian in Fock space with a small variance $\langle \! \langle n^2 \rangle \! \rangle \ll \bar n$ for large pumping strengths $\epsilon \gg \gamma_a$. As such, the stationary state of the system is very close to a Fock state and exhibits the same features. In particular, the Wigner function is negative as shown in the main text.
\end{quote}
\subsection{Fock state stabilization}
Here, we want to solve the Lindblad equation
\begin{equation}\label{eq:Lindblad_S}
\dot \rho = \gamma_a\mathcal{D}[a]\rho + \epsilon \mathcal{D}[a^\dag \, {}_1\!F_1(-a^\dag a, 2, \alpha_a)]\rho
\end{equation}
for the stationary state where $\dot\rho=0$.
Numerically, we find that the stationary probability distribution $P_n=\langle n| \rho | n \rangle$ is well described by a Gaussian distribution $P_n \propto \exp[(n-\bar n)^2/2\sigma^2]$ with mean $\bar n$ and variance $\sigma^2 = \langle \! \langle n^2 \rangle \! \rangle$.
Thus, the goal is to find a diffusion equation in the form of $\partial_n [A(n) + \partial_n B(n)]P_n=0$ with drift and diffusion coefficients $A(n)$ and $B(n)$.
In particular, for $A(n) \approx A_0(n-\bar n)$ with $A_0$ and $B(n) \approx B_0$ being constant, the solution is given by $P_n \propto \exp[A_0(n-\bar n)^2/2 B_0]$.
Thus, the variance will be given by $B_0/A_0$ and the average photon number is determined by $A(\bar n)=0$.

From Eq.~\eqref{eq:Lindblad_S}, it is possible to derive a rate equation for the probabilities $P_n$ given by 
\begin{equation}\label{eq:rate_equation}
\dot P_n = \gamma_a \bigl[(n+1)P_{n+1} - n P_n\bigr] + \epsilon \bigl[n f(n-1)^2 P_{n-1} - (n+1)f(n)^2 P_n\bigr]
\end{equation}
with $f(n) =  {}_1\!F_1(-n, 2, \alpha_a)$.
While this is still exact, it is not possible to solve this rate equation due to the nonlinearity $f(n)$. 
However, we will make a set of approximations in the following. 
First, since we are interested in Fock states with a large photon number $n_0 \gg 1$, we can employ a Kramers-Moyal expansion up to second order with 
\begin{equation}
P_{n\pm 1} \approx P_n\pm \partial_n P_n + \frac12 \partial_n^2 P_n.
\end{equation}
Then, we are able to cast Eq.~\eqref{eq:rate_equation} into a diffusion equation as stated above with drift and diffusion coefficients given by
\begin{equation}
A(n) = n \gamma_a - \epsilon (n+1) f(n-1)\bigl[f(n-1) - 2 \partial_n f(n-1)\bigr] ,
\end{equation}
\begin{equation}
B(n) = \frac12\bigl[(n+1)\gamma_a + \epsilon \, n f(n-1)^2\bigr].
\end{equation}

First, we want to obtain the average photon number $\bar n$ from the drift coefficient $A(n)$.
As explained above, the average photon number is characterized by $A(\bar n)=0$.
Again, due to the highly nonlinear behavior of $f(n)$, it is not possible to solve this in general.
However, we are interested in the case of strong pumping with $\epsilon \gg \gamma_a$ where the stationary state is close to the Fock state $n_0 \gg 1$. As such, $\bar n$ will only have a small deviation to $n_0$ which we need to determine.

Due to the blockade mechanism, which is explained in the main text, we are able to linearize $f(n) \approx f_0 (n-n_0)$ where $f_0$ determines the slope at $n_0$.
To determine $f_0$, we make use of the Mehler-Heine formula
\begin{equation}\label{eq:Mehler}
f(n) = {}_1\!F_1(-n, 2, \alpha_a) \approx e^{\alpha_a/2} \frac{J_1(\sqrt{4n\alpha_a})}{\sqrt{n\alpha_a}}
\end{equation}
which is asymptotically valid for $n\gg1$, introducing the Bessel function of the first kind $J_m(x)$.
Thus, the root of $f(n)$ is given by the root of the Bessel function $J_1(\sqrt{4n\alpha_a})$.
The first root of the Bessel function is given by $x_{1,1} \approx 3.83$. 
Thus, to stabilize the Fock state with photon number $n_0\gg1$, $\alpha_a$ has to be (approximately) chosen as $\alpha_a = x_{1.1}^2/4n_0 \approx 3.67/n_0$.
Since $\alpha_a$ is small for large $n_0$, we can safely discard the exponential prefactor in~\eqref{eq:Mehler}.

The slope $f_0$ can be obtained from the derivative of Eq.~\eqref{eq:Mehler}.
It is given by
\begin{equation}
f'(n) \approx \frac{J_0(\sqrt{4n\alpha_a})}{n} - \frac{J_1(\sqrt{4n\alpha_a})}{\sqrt{4n^3\alpha_a}}.
\end{equation}
The second term drops out since we evaluate the derivative at the root of $J_1(x)$.
Thus, the slope is given by $f_0 = J_0(x_{1,1})/n_0 \equiv J/n_0$ with $J \approx -0.40$ as stated in the main text.

Using the linearized form $f(n)\approx J(n-n_0)/n_0$, we are able to solve $A(\bar n) = 0$.
For $n_0 \gg 1$, we obtain
\begin{equation}
\bar n = n_0 \frac{\sqrt{\gamma_a+ 4J^2 \epsilon} - \sqrt{\gamma_a} }{4J^2 \epsilon} \approx n_0\Bigl(1-\sqrt{\frac{\gamma_a}{J^2\epsilon}}\Bigr).
\end{equation}

Next, we want to obtain the variance $\langle \!\langle n^2\rangle \! \rangle$.
For this purpose, we linearize the drift $A(n) \approx A_0 (n-\bar n)$ and assume the diffusion to be constant with $B(n)\approx B(\bar n)=B_0$.
As stated above, the variance is then given by $B_0/A_0$ where $A_0 = \sqrt{4J^2\gamma_a\epsilon} + \gamma_a$ and $B_0= n_0(\gamma_a+\sqrt{\gamma_a^3/J^2\epsilon})$ for $n_0 \gg 1$.
This leads to the variance
\begin{equation}
\langle \! \langle n^2 \rangle \! \rangle = n_0\frac{\sqrt{J^2 \gamma_a \epsilon}-\sqrt{\gamma_a}}{2J^2\epsilon +\sqrt{J^2\gamma_a\epsilon}} \approx n_0\sqrt{\frac{\gamma_a}{4J^2\epsilon}}
\end{equation}
as given in the main text.
Thus, the probability distribution $P_n$ is a narrow Gaussian and the Fock state is reached for strong pumping with $\epsilon^{-1/2}$.

\subsection*{Implementation in QuTiP}
In this section, we show how to define the deformed ladder operator in QuTiP and show, as an example, how to implement the single photon loss dissipator.
\begin{lstlisting}[language=Python,frame=single]
from qutip import *

# Implementation of the deformed ladder operators using QuTiP
#
# Parameter:   q (float) : counting field 

M = 10 # Hilbert space truncation

# deformed ladder operator
def deformed_supop(q):
    ap = spre(destroy(M, offset=q))    # (deformed) superoperator acting from the left
    am = spost(destroy(M, offset=0))   # superoperator acting from the right
    apd = ap.dag()
    amd = am.dag()
    
    return ap, apd, am, amd

#Example: Single photon loss dissipator

def single_ph_loss(q):
    ap, apd, am, amd = deformed_supop(q)
    return ap*amd - apd*ap/2 - am*amd/2
\end{lstlisting}

\end{document}